\documentclass[aps,shownopacs,superscriptaddress,nofootinbib,preprint,11pt]{revtex4}
\usepackage{hyperref}
\usepackage{amsmath}
\usepackage{float}
\usepackage{graphics}
\usepackage{graphicx}
\usepackage{epsfig}
\usepackage{psfrag}
\usepackage{color}
\usepackage{slashed}

\usepackage{amsfonts}
\usepackage{amssymb}

\newcommand*{\be}{\begin{equation}}
\newcommand*{\ee}{\end{equation}}
\newcommand*{\bea}{\begin{eqnarray}}
\newcommand*{\eea}{\end{eqnarray}}
\newcommand*{\e}{\epsilon}

\newcommand{\comment}[1]{}

\newcommand{\cref}[1]{Chapter~\ref{c.#1}}

\def\beq{\begin{equation}}
\def\eeq{\end{equation}}
\def\bea{\begin{eqnarray}}
\def\eea{\end{eqnarray}}
\def\ba{\begin{array}}
\def\ea{\end{array}}
\def\bi{\begin{itemize}}
\def\ei{\end{itemize}}
\def\be{\begin{enumerate}}
\def\ee{\end{enumerate}}
\def\bc{\begin{center}}
\def\ec{\end{center}}
\def\bt{\begin{table}}
\def\et{\end{table}}
\def\btb{\begin{tabular}}
\def\etb{\end{tabular}}

\def\lsim{\raise0.3ex\hbox{$\;<$\kern-0.75em\raise-1.1ex\hbox{$\sim\;$}}}
\def\gsim{\raise0.3ex\hbox{$\;>$\kern-0.75em\raise-1.1ex\hbox{$\sim\;$}}}

\begin{document}

\title{Warped Alternatives to Froggatt-Nielsen Models}
\author{Abhishek M Iyer}
\email{abhishek@cts.iisc.ernet.in}
\affiliation{Centre for High Energy Physics, Indian Institute of Science,
Bangalore 560012}
\author{ Sudhir K  Vempati}
\email{vempati@cts.iisc.ernet.in}
\affiliation{Centre for High Energy Physics, Indian Institute of Science,
Bangalore 560012}

\begin{abstract}
We consider the Randall-Sundrum (RS) set-up to be a theory of flavour, as an alternative to Froggatt-Nielsen (FN) models
instead of as a solution to the hierarchy problem. The RS framework is modified by taking the low energy brane to be at the GUT scale.
This also alleviates constraints from  flavour
physics. Fermion masses 
and mixing angles are fit at the GUT scale. 
The ranges of the bulk mass parameters  are determined using a $\chi^2$ fit taking in to 
consideration the variation in $\mathcal{O}(1)$ parameters. In the hadronic sector, 
the heavy top quark requires large bulk mass parameters localising the right handed
top quark close to the IR brane.  Two cases of neutrino 
masses are considered (a) Planck scale lepton number violation and (b) Dirac neutrino masses. 
Contrary to the case of weak scale RS models, both  these cases give reasonable fits to the data, 
with the  Planck scale lepton number violation fitting slightly better compared to the Dirac case. 
In the Supersymmetric version, the fits are not significantly different except for the variation in $\tan\beta$. 
If the Higgs superfields and the SUSY breaking spurion are localized on the same brane then the structure
of the sfermion masses are determined by the  profiles of the zero modes of the hypermultiplets in the bulk. 
Trilinear terms have the same structure as the Yukawa matrices.  The resultant squark  spectrum is around $\sim 2-3~ \text{TeV}$ 
required by the light Higgs mass to be around 125 GeV and  to satisfy the flavour violating constraints. 

\end{abstract}
\vskip .5 true cm

\pacs{73.21.Hb, 73.21.La, 73.50.Bk}
\maketitle
\section{Introduction}
One of the celebrated solutions of the fermion flavour problem is the Froggatt-Nielsen Mechanism \cite{FN}. According to this prescription, the symmetry group of the Standard Model (SM) is augmented by a horizontal
$U(1)_X$ group under which all the SM fermions and the Higgs field are charged. The effective theory includes a flavon field $X$ and the Yukawa couplings are generated from the higher dimensional operators which 
are invariant under the $U(1)_X$ and the Standard Model (SM) gauge group. For example, the up-type quark mass matrix
has the form:  $Y^u_{ij} (\frac{X}{M_{Pl}})^{c_{Q_i}+c_{u_j}+c_{H_u}}Q_iH_u U_j$, 
where $c_{f}$ is $U(1)_X$ charge of the  $f$
field and $i,j$ are the generation indices. The flavon field $X$ develops a vacuum expectation value (vev) 
 such that $0.22 \approx \lambda_c \approx  <X> /M_{Pl}$, $\lambda_c$ being the Cabibbo angle. 
$Y^u_{ij}$ are taken to be  $\mathcal{O}(1)$ parameters. Fermion mass matrices including their mixing
patterns can be fit to the data by choosing appropriate $U(1)_X$ charges for various fields.  An UV completion
of the model can be constructed by including heavy chiral fermions in to the theory; integrating these heavy
fields would lead to the relevant non-renormalizable operators (for a review, see \cite{Babu} ). 
   
 The $U(1)_X$ symmetry introduces additional anomalies in to the theory and subsequently, strong constraints
 on the $U(1)_X$ charges for various fields. In supersymmetric models with a single flavon field, one typically  
 has to resort to Green-Schwarz (GS) mechanism to cancel the anomalies. 
    The solution set of $U(1)_X$ charges for the fermions and the Higgs which satisfy the fermion data as well as the anomaly cancellation\footnote{ However, with two singlet flavons there exist a unique solution which is completely non anomalous \cite{Dudas:U(1)}} requirement 
    have been studied in \cite{Dudas:U(1),ibarra,king,Ibanezross,Binetruy1,Binetruy2,Chun,Dreiner1,Dreiner2,King:2004tx,Ellis,Joshipura:2000sn} and recently updated
    in \cite{lavignac}.  These models typically lead to large flavour violations at the weak scale in gravity mediated supersymmetry breaking models due to contributions from the $U(1)_X$ D-terms. 
        While the constraints from the flavour sector on the available solutions are very tight, it may still be possible to ease them without requiring the superpartner masses to be very high \cite{Lalak,Varzielas}. 
The flavour constraints may also be alleviated to some extent by considering $U(1)\times U'(1)$ class of models \cite{Leurer,flavour}. In the present work, we will study the extra-dimensional alternative \cite{ArkaniHamed} to 
understand the flavour hierarchy in particular concentrating on the supersymmteric Randall-Sundrum (RS) set up.

   The Randall-Sundrum framework \cite{RS} which elegantly provides a solution to the hierarchy problem via warping in the
   extra dimensional space can also thought to be a theory of flavour. 
   It has been observed sometime ago that the flavour changing neutral currents (FCNC)  can be suppressed due to the so-called 
   RS-GIM mechanism \cite{RSGIM}. However, in the absence of additional flavour
    symmetries the constraints from FCNC are still very strong(\cite{Huber,Agashe,Delaunay,Petriello,AgasheSundrum}) (Detailed analysis for the
    hadronic sector can be found in \cite{Neubert1,Neubert2} and references there in. For a recent thorough analysis
    in the leptonic sector, please see \cite{iyer}).  Given these strong constraints on the RS set up at the weak scale,  one can
    ask the question whether RS is suitable to be a theory of flavour as well as a solution to the hierarchy problem simultaneously. 
       It might be that RS as a theory of a flavour might be better suited at the GUT scale rather than at the weak scale. 
    The Froggatt-Nielsen models are typically defined at scales closer to the Planck scale, so perhaps flavour physics might 
    have its origins at the Planck scale. 

  With this rationale,  in the present work we will consider RS to span between the Planck and the GUT scales. 
     The fermion masses are fit in terms of the bulk mass parameters of the various fields, which take the role of the $U(1)_X$
charges of the FN mechanism. However, these parameters are less constrained  compared to the $U(1)_X$ charges, as no
additional conditions such as anomaly cancellations are required on them. 
 While this has been the common understanding, in \cite{Dudas} it was pointed out that imposing unification conditions on gauge couplings
 in a theory with localization of fermions or hierarchical wave functions leads to strong constraints which are exactly in the same
  way as the Green-Schwarz anomaly cancellation conditions \cite{Greenschwarz} \footnote{Typically applied in FN models, the Green-Schwarz anomaly
  cancellation conditions requires the anomaly factors to be in a particular ratio such that they are cancelled in String theory.} 
  In the present setup we do not impose these conditions. 
    
  Extra dimensions at GUT scale were considered in \cite{Hallnomura} while the RS version was considered by the authors in \cite{choi1,choi2} and later by the authors in \cite{Dudas}. Our work, however,  is very closely related  to the work of \cite{Brummer} who have done a thorough analysis of fermion mass spectrum,  weak scale supersymmetric spectrum and flavour phenomenology, assuming a particular
 Grand Unified Theory (GUT)  model in such a RS setting. However, differences exist.  
 In the present work we have not  assumed any specific GUT model.  Furthermore, we have used a frequentist approach to do the  fermion mass fitting.  While this makes it hard to directly compare   the results between the two works, we hope they provide a complementary set of results.    We also have taken in to consideration  the constraints from neutrino masses 
  and mixing angles  which can have a significant effect on the lepton flavour violation and slepton decays.
  
   The equivalent description for the RS set up in four dimensions can be thought of as a composite Higgs 
 coupled to fermions with couplings which parameterise the  \textquoteleft partial compositeness\textquoteright \hspace{0.06cm}  of the fermions\cite{Rattazzi}. 
 In SUSY case, this partial compositeness can also affect the structure of the soft masses.
  
    In the first part of our work,  our aim has been to provide a range of bulk mass parameters which fit the fermion masses 
    and mixing patterns at the GUT scale.
 We believe this can be useful for model builders and other phenomenologists working in flavour physics  and looking for an
 alternative to FN models.   
   We have considered both supersymmetric as well as non supersymmetric versions of the RS framework at the GUT scale 
   while fitting the data. The supersymmetric case has the added advantage that it could lead to observable 
   signatures at the weak scale.  We consider the case where  SUSY breaking
   is considered to be on the same brane as where the Higgs is localized, which is the GUT brane.
     In this case,  the sfermion mass matrices are determined by the zero mode profiles of the corresponding N=1 superfields
      and thus the information of the fermion masses is propagated in to the soft sector. It is far more striking for the A-terms
      which follow the same structure as the Yukawa couplings. The spectrum  is highly non-universal at the high scale, but, 
      its pattern is constrained due to the ranges of bulk mass parameters which are in turn are fixed by their fits to fermion masses. 
     The running effects make the diagonal terms large at the weak scale.
     
        The rest of the paper is organized as follows. In section \ref{rs-setup}, 
        we detail the RS setup we consider and derive the structure of the
        fermion masses.  In section III we present the fermion mass fits and present the ranges for the bulk mass parameters for
        both the non-supersymmetric and the supersymmetric cases.
   In section IV we address the issue of supersymmetric breaking and derive supersymmetric spectrum for a particular
   supersymmetric breaking case.  We end with summary and outlook in the last section.   In Appendices \textbf{A} , \textbf{B}and $\textbf{C}$,
   we have presented plots relevant for fermion mass fits.

\section{RS as a theory of flavour}
\label{rs-setup}

The Randall-Sundrum frame work consists of two branes separated by an single warped extra dimension \cite{RS}.
The line element for the RS background is given as 
\begin{equation}
 ds^2=e^{-2ky}\eta_{\mu\nu}dx^\mu dx^\nu-dy^2
\end{equation}
where $0\leq y\leq\pi R$. Here $y=0$ is identified as the position of the UV brane and $y=\pi R$ is the position of the IR brane. 
The scale associated with physics on the UV brane is $M_{Pl}$ while that on the IR brane is $\text{TeV}$. The solution to the
 hierarchy problem is achieved by exponential warping of scales i.e, $M_{Pl}=e^{kR\pi} ~M_\text{weak}$, where
  $kR\sim\mathcal{O}(11)$. 
  
  In the modified set up we consider here,  the scale associated with the IR brane is $M_{GUT}$.
This can be achieved by choosing $kR \sim 1.5$. We define the hierarchy between the scales,  $\epsilon$, for this scenario to be 
\begin{equation}
 \epsilon=\frac{M_{GUT}}{M_{Planck}}\sim 10^{-2}
\label{warpfactor}
\end{equation}
We consider both supersymmetric and non-supersymmetric matter fields to propagate in the bulk. Localisation of the 
respective zero modes is dependent on the corresponding bulk masses. In both the cases, we assume that the Higgs
field (two Higgs fields  in the case of supersymmetric models)  are localised on the GUT brane.
We now proceed to briefly review the derivation of the mass matrices and their dependence on the zero mode profiles
in both supersymmetric and non-supersymmetric cases. 
A couple of points are important to note at this juncture. Firstly,  the typical lowest KK mass for a warped background is given as
$m_{KK}=e^{-kR\pi}k$.  In the present set up, the  \textquoteleft large\textquoteright \hspace{0.06cm} warp factor ensures the lowest KK modes are very heavy i.e, 
$m_{KK}=\epsilon k \sim M_{GUT}$ and thus are decoupled from low energy phenomenology. 
We do not consider their effects in this work for low energy phenomenology.  Secondly, it turns out that  the dependence of the 
zero mode  mass matrices on the profiles is very similar in supersymmetric and non-supersymmetric cases. However, the 
fermion mass data at the high scale in the supersymmetric case could be different from the SM one, due to the dependence
on tan$\beta$ as well as the different RGE for the Yukawa coupling as we will discuss in the next section.  

\subsection{Standard Model case} 

The non-supersymmetric case or the Standard Model case has been studied in many works \cite{Neubert1,Neubert2,Huber,iyer}. 
The main difference in the present case is that while those studies have considered a $kR \sim  \mathcal{O}(11)$, while in the present case
it is $\mathcal{O}(1)$.  We thus assume a grand desert from the weak scale to the GUT scale, where RS framework sets in.
No attempt is made to solve the hierarchy problem, but the flavour problem has a solution in terms of the localisation of the fields
in an extra dimension at the GUT scale. We follow the notation of  \cite{iyer} and present the final formulae for the Yukawa
mass matrices. The details of the KK expansion  and the corresponding ortho-normal relations can be found in \cite{iyer} and references therein. 
The five dimensional action has the form: 
\begin{eqnarray}
\label{smaction}
S &=& S_{\text{kin}} + S_{\text{Yuk}} + S_{\nu} + S_{higgs} \nonumber \\ 
 S_{kin} &=& \int d^4x\int dy ~\sqrt{-g}~ \left(~ \bar{L} (i\slashed D - m_L)L + \bar{E} (i\slashed D - m_{E})E  + \ldots ~\right) \nonumber \\ 
S_{\text{Yuk}}  & =& \int d^4x\int dy~\sqrt{-g} \left(  ~ Y_U \bar{Q}U \tilde{H} +Y_D \bar{Q}DH
+ Y_E \bar{L} E H \right) \delta(y-\pi R)  \nonumber \\ 
S_{\nu} &= &  \int d^4x\int dy~\sqrt{-g} \left(  \frac{\mathbf{\kappa}}{\Lambda^{(5)}}LHLH ~~ \text{\textbf{or}}~~ 
Y_N \bar{L} N H \right) \delta(y-\pi R) \end{eqnarray}
where we used the standard notation with the $Q,U,D$  standing for the quark doublets, up-type and down type singles respectively, $L$
and $E, N$ stand for leptonic doublets and charged and neutral singlets respectively. $H$ stands for the Higgs doublet with $\tilde{H} 
= i \sigma H^\star$. We have suppressed the Higgs action in the above. Two specific ways for generating non-zero
neutrino masses are considered (a) by a higher dimensional term localised at the GUT brane and (b) Dirac neutrino mass terms similar to the other
fermions. $\Lambda^{(5)}$ is the five dimensional reduced Planck scale $\sim 2 \times 10^{18}$ $\text{GeV}$. 

After the Kaluza-Klein (KK)  reduction and imposing the orthonormal conditions, we can derive the 4D mass matrices for the zero-modes 
of the fermion fields.  They have the form: 
\begin{eqnarray}
\label{SMmasses}
 ({\mathcal M}_{F})_{ij} &=&\frac{v}{\sqrt{2}} ({Y}_F')_{ij}  e^{(1-c_i-c'_j)k R \pi} ~\xi(c_i) ~\xi(c'_j) \;\;\;;\;\;
 \nonumber \\ 
 \xi(c_i) &=&  \sqrt{\frac{(0.5-c_{i}) }{e^{(1-2 c_i)\pi k R}-1}},
 \end{eqnarray}
 where $F$ stands for all the Yukawa matrices $F = U, D, E $ and $N$, if the neutrinos have Dirac masses. $c_i$ and $c'_j$ represent
 the bulk masses of the respective matter fields (second line of eq.(\ref{smaction}));  
 defined as, for example,  $m_{E_i} = c_{E_i} k$. $i,j$  denote the generation indices. If the neutrinos have Dirac
 masses then their mass matrix is given by Eq.(\ref{SMmasses}).
 In case the neutrinos attain their masses
 through higher dimensional operator, the mass matrix is given by
 \begin{equation}
 \label{smneutrino1}
({\mathcal M}_\nu)_{ij} =\frac{v^2 }{2\Lambda^{(5)}} (\kappa')_{ij} e^{(2-   c_{L_i}-c_{L_j}) kR\pi }  \xi(c_{L_i}) \xi (c_{L_j})
\end{equation}
 In Eqs. (\ref{SMmasses}, \ref{smneutrino1}), we have defined   $Y' = kY$  and $\kappa'=2k\kappa$. These are  dimensionless 
 $\mathcal{O}(1)$ parameters of anarchical nature\footnote{ Note that the Yukawa couplings in 
Eq.(\ref{superpotential-yuk}) are dimensionful, with mass dimensions -1.}.  Eq.(\ref{SMmasses}) are used to fit  all the fermion mass data at the GUT scale 
 \textit{i.e,} up and down type quark masses and the  (Cabibbo-Kobayashi-Masakawa) CKM matrix, charged lepton masses, neutrino mass differences and the corresponding PMNS mixing matrix.  In the case neutrinos get their masses through higher dimensional operator,
 Eq. (\ref{smneutrino1}) is used instead to fit their mass differences and mixing angles.

 \subsection{Supersymmetric case} 
 
 In the supersymmetric case, the matter  fermions are represented  by hyper-multiplets\footnote{ N=1 Supersymmetry in 5D has the particle
 content of N=2 Supersymmetry in 4D. The hypers can be expressed in N=1, 4D language as two chiral superfields, 
 where as the Vectors can be expressed as a vector and chiral superfield \cite{Marti,ArkaniHamed:2001tb}.} 
 propagating in the bulk. In terms of the 4D, N=1 SUSY  language, they can be expressed as two N=1 chiral multiplets, 
 $ \Phi, \Phi^c$.  Following \cite{Marti,Gherghetta1}, we write the 5D action in terms of two chiral fields with a (supersymmetric) bulk 
 mass term to be
\begin{equation}
 S_5=\int d^5 x\left[\int d^4\theta e^{-2ky}\left(\Phi^\dagger\Phi+\Phi^c\Phi^{c\dagger}\right) + \int d^2\theta e^{-3ky}\Phi^c\left(\partial_y+M_\Phi-\frac{3}{2}k\right)\Phi\right]
\label{chiralsuperfield}
\end{equation}
where $M_\Phi =c_\Phi k$ is the bulk mass.
In writing the above, the radion field is suppressed by taking  its vacuum expectation value,  $<Re(T)>= R $.  
 The super field  $\Phi^c$ is taken to be odd under $Z_2$.  Thus, only $\Phi$ has a zero mode. 
 Since we have a theory at the GUT scale, the KK modes can be considered to be decoupled from theory. In the 
 effective theory, the profile of the zero mode of the $\Phi$ is determined by\cite{Marti}
\begin{equation}
 \left(\partial_y-\left(\frac{3}{2}-c\right)k\right)f^{(0)}=0
 \label{profile}
\end{equation}
Thus $f^{(0)}=  e^{(\frac{3}{2}-c)ky}$.
The superscript $^{(0)}$ stands for the zero mode, which we will drop subsequently\footnote{In the
component form, the scalar component and the fermion components of the chiral super field $\Phi$ have different bulk masses.
However, the solution for the profile for the scalar and the fermion components turns out to be the same. }.   
In this effective theory, where the higher KK modes are completely decoupled, 
 we can write the effective 4-D K{\"a}hler terms for the $Z_2$ even zero modes as \cite{Dudas,choi1,choi2}
\begin{eqnarray}
\mathcal{K}^{(4)}&=& \int dy\left(
e^{(1-2c_{q_i})k y }Q^\dagger_iQ_i+
e^{(1-2c_{u_i})k y }U^\dagger_iU_i+
e^{(1-2c_{d_i})k y }D^\dagger_iD_i + \ldots \right),
\label{kahler1}
\end{eqnarray}
where we have substituted for the  profile solutions of (\ref{profile}). 
After integrating over the extra dimension $y$, the  terms in Eq.(\ref{kahler1}) pick up a factor $Z_{F}=\frac{1}{(1-2c_F)k}\left(
\e^{2c_F-1}-1 \right) $ where $F=Q,U,D, L, E$, as before. We choose to work in a basis in which the K{\"a}hler terms are 
canonically normalized. We thus re-define the fields as $\Phi\rightarrow \frac{1}{\sqrt{Z_F}}\Phi$. 

The effective four dimensional MSSM Yukawa couplings are determined from the superpotential terms written on the boundary.
For Higgs localized on the IR brane, keeping only the zero modes of the chiral superfields, the effective four dimensional superpotential is given as \cite{Dudas,Marti}
\begin{eqnarray}
 \mathcal{W}^{(4)}&=&\int dy e^{-3ky}\left(e^{(\frac{3}{2}-c_{q_i})ky}e^{(\frac{3}{2}-c_{u_j})ky}Y^{u}_{ij}H_UQ_iU_j+e^{(\frac{3}{2}-c_{q_i})ky}e^{(\frac{3}{2}-c_{d_j})ky}Y^{d}_{ij}H_DQ_iD_j\right.\nonumber\\
 &+&  \left. e^{(\frac{3}{2}-c_{L_i})ky}e^{(\frac{3}{2}-c_{E_j})ky}Y^{E}_{ij}H_DL_iE_j  + \ldots  \right) \delta(y-\pi R)
 \label{superpotential-yuk}
\end{eqnarray}
The Higgs fields are canonically normalized as $H_{u,d}\rightarrow e^{kR\pi}H_{u,d}$. In the canonical
basis, after the fields have been redefined the fermion mass matrices can be derived from Eq. (\ref{superpotential-yuk}) to be 
\begin{eqnarray}
 (\mathcal{M}_F)_{ij}=\frac{v_{u,d}}{\sqrt{2}}Y'_{ij}e^{(1-c_i-c'_j)kr\pi}  \xi (c_i) \xi (c'_j)
 \label{chargedfermionIR}
 \end{eqnarray}
 where $c_i,c'_j$ denote the bulk mass parameters for various fields. $\xi(c_i)$ are defined in Eq.(\ref{SMmasses}) 
 The mass matrix in Eq.(\ref{chargedfermionIR}) can be approximated as  $(\mathcal{M}_F)_{ij}\sim \frac{v_{u,d}}{\sqrt{2}}\mathcal{O}(1)e^{(1-c_i-c'_j)kr\pi}$
 where the $\xi(c)$ is absorbed into the $\mathcal{O}$(1) parameters $Y'$ and is now collectively referred to as $\mathcal{O}$(1).
 This is true only as long as the $c$ parameter lies between 0 and 1. But as we have seen earlier, in some realistic
 cases especially related to neutrino masses and the top quark, the values of $|c|$ could be large to fit the data. Redefining the $\mathcal{O}$(1) Yukawa by absorbing the c parameters
 would shift the ranges of the $c$ parameters far away from what they are, especially in the case where $|c|\geq 1$.
As in the SM case, we define dimensionless $\mathcal{O}(1)$ Yukawa couplings as $Y'=2kY$ and are of anarchical
nature. While the Dirac masses for the neutrinos have the same structure as the other fermion mass matrices, the higher dimensional
operator has a different form determined by the super-potential term 
\begin{eqnarray}
 \mathcal{W}^{(4)}=\int dy \delta(y-\pi R)e^{-3\sigma(y)}\left(e^{(\frac{3}{2}-c_{L_i})ky}e^{(\frac{3}{2}-c_{L_j})ky}\frac{\kappa_{ij}}{\Lambda^{(5)}}H_UH_UL_iL_j\right)\nonumber\\
 \label{superpotential11}
\end{eqnarray}
The neutrino mass matrix in this case is given as 
\begin{equation}
  (\mathcal{M}_\nu)_{ij} = \kappa'_{ij}\frac{v_u^2sin^2(\beta)}{2\Lambda^{(5)}} e^{(2 -c_{L_i}-c_{L_j})kR\pi} \xi(c_{L_i}) \xi(c_{L_j})
  \label{neutrinomassmatrix1}
 \end{equation}
where $\kappa'=2k\kappa$ is the dimensionless $\mathcal{O}$(1) parameters. The fermion mass matrices carry the same form as in the SM and the supersymmetric cases and thus their
dependence on $c_i$ and $\mathcal{O}(1)$ parameters is the same.

\section{Fermion Mass fits}
From the previous section, we have seen that in addition to the bulk mass parameters,  the $\mathcal{O}(1)$ Yukawa parameters also play a role in fixing the fermion masses and mixing angles. 
We fit the masses and the mixing angles  of the quark sector and the neutrino sector
 at the GUT scale for both  the SM and the supersymmetric cases.  We will use a frequentist 
 approach, \textit{i.e,} we minimise the $\chi^2$ function, which is defined as follows: 
\begin{equation}
 \chi^2=\sum_{j=1}^N\left(\frac{y_j^{exp}-y_j^{theory}}{\sigma_j}\right)^2
 \label{chisq}
\end{equation}
where, $y_j^{theory}$ is the theory number for the $j^{th}$ observable and $y_j^{exp}$ is its corresponding number quoted by experiments with a measurement uncertainty of $\sigma_j$. In the present case the theory parameters 
are just the bulk mass parameters and the $\mathcal{O}(1)$ Yukawa entries in supersymmetric and non-supersymmetric
cases.  We define $0<\chi^2<10$ to be a good fit and we try to find regions in the parameters space of bulk mass parameters
and $\mathcal{O}(1)$ Yukawa parameters which satisfy this condition\footnote{We will mention the results with lower
$\chi^2$ at relevant places.}. 
The  $\mathcal{O}(1)$ Yukawa parameters are varied between -4 and 4, 
with a lower bound of 0.08 on $|Y|$ to avoid unnaturally small Yukawa parameters. As far as the bulk mass parameters
are concerned, since they are given as $c k$,  we prefer to vary the $c$ parameters between  $-1$ to $1$, so as  not to
go beyond the 5D cut-off, $k$.   This will remove any possible inconsistencies in the theory due to non-perturbative Yukawa couplings. 
However, as we will see it is not always possible to fit the data within this range of $c$ parameters.  We will mention the
range chosen specifically for each case. 

The minimization of the $\chi^2$ function
was performed using MINUIT \cite{minuit}. We can minimise the hadronic and the
leptonic sectors independently as they are dependent on different  sets of parameters which are uncorrelated. 
The methodology is similar to the ones used in fermion mass fitting
in GUT models \cite{Joshipura,Altarelli} and also the one used in \cite{iyer}.       

\subsection{Standard Model (SM) Case}
 In this section, we present  the fits in the SM case.  For the GUT scale values of the quark and
 lepton masses and CKM mixing matrices,  we use the results of  \cite{xing}. 
 In the analysis of \cite{xing},  two loop RGE have been used to run the Yukawa couplings of the 
 up-type quarks, down-type quarks  and charged leptons  from the weak   scale all the way  up to the GUT scale. 
 For the neutrino data we used the
 publicly available package REAP \cite{reap} to compute  the high scale values. The masses of the SM 
 fermions at the GUT scale used in our fits are presented  in Table \ref{inputtable1}.  The CKM and PMNS 
 mixing matrices are presented in Table \ref{inputtable2}. 

 \begin{table}[htbp]
\caption{GUT scale masses of fermions for the SM case}
\begin{center}
\begin{tabular}{|c|c|c|c|}
\hline 
Mass &Mass&Mass&Mass squared Differences\\
(MeV )& (GeV)& MeV&$eV^2$ \\
\hline
\hline
$m_u = 0.48^{+0.20}_{-0.17}$&$m_c=0.235^{+0.035}_{-0.034}$&$m_e = 0.4696^{+0.00000004}_{-0.00000004}$&$\Delta m^2_{12} = 1.5  ^{+0.20}_{-0.21} \times 10^{-4} $\\
$m_{d} = 1.14^{+0.51}_{-0.48}$&$m_b=1.0^{+0.04}_{-0.04}$&$m_{\mu} = 99.14^{+0.000008}_{-0.0000089}$&$\Delta m_{23}^2 = 4.6^{+0.13}_{-0.13}  \times 10^{-3}    $\\
$m_{s} = 22^{+7}_{-6}$&$m_t=74.0^{+4.0}_{-3.7}$&$m_{\tau} = 1685.58^{+0.19}_{-0.19}$&-\\
\hline
\end{tabular}
\end{center}
\label{inputtable1}
\end{table}

\begin{table}[htbp]
\caption{Mixing angles for the hadronic and the leptonic sector for the SM case}
\begin{center}
\begin{tabular}{|c|c|c|}
\hline 
mixing angles(CKM)&Mixing angles (PMNS)\\
\hline
\hline
$ \theta_{12} =0.226^{+0.00087}_{-0.00087}$ &$ \theta_{12} =0.59^{+0.02}_{-0.015}$ \\ 
$\theta_{23} = 0.0415^{+0.00019}_{-0.00019}$&$\theta_{23} = 0.79^{+0.12}_{-0.12}$\\
$\theta_{13}=0.0035^{+0.001}_{-0.001}$&$\theta_{13}=0.154^{+0.016}_{-0.016}$\\
\hline
\end{tabular}
\end{center}
\label{inputtable2}
\end{table}
     
\subsubsection{SM Quark Sector  fits}
The up and down mass matrices are given in terms of fermion mass matrix 
of  Eq.(\ref{SMmasses}). The theory parameters  which are varied simultaneously to minimise 
the $\chi^2$ in Eq.(\ref{chisq}) include:  three $c_{Q_i}$, each of $c_{u_i}$ and $c_{d_i} $  and 18 $\mathcal{O}(1)$ 
Yukawa  parameters.  We would expect that the light quarks would be localised close to the UV brane ( $c > 1/2$ ) 
and the heavy quarks close to the IR brane ($c <  1/2$). However, for this particular range of $\mathcal{O}(1)$ Yukawa
parameters, it is difficult to fit the data for $|c|$ within unity. We thus enlarged the range for the $c$ parameters. 

The range chosen for the scan of the $c$ parameters chosen is: $-2<c_{Q_1,Q_2}<4$, $-3<c_{Q_3}<1$ for the doublets.
$-2<c_{d_1,d_2,d_3}<3.5$, for the down type singlets and $-2<c_{u_1,u_2}<4$, $-4<c_{u_3}<1$ for the up type singlets. 
 We  fit  the quark masses and the CKM mixing angles at the GUT scale.  
 The top quark is definitely lighter at the GUT scale,
 but still we see that most of the points that fit the data lie outside of $|c|~ \leq ~1$. 
This is evident from the the negative values of the $c_{Q_3}$ and $c_{U_3}$ that fit the data.

The regions of $c$ parameter space which satisfy the  constraint of  $0<\chi^2<10$ for the chosen scanning range are shown
 in Fig.(\ref{quarksm}) in Appendix A and the ranges are outlined in Table[\ref{hadronrangesm}].  
We see that the first two generation bulk mass parameters  are concentrated on the positive $c$ values where as
the third generation, the doublet and more so  the right handed top is localised close to the GUT scale brane. 
Comparing these results with that of the normal RS, we find that the masses for the light quark fields can be fit with 
$c\sim 0.6-0.7$. This can be attributed to the large warping where $0.5<c<1$ is sufficient to reproduce the masses for light quarks \cite{Hubershafi}.

\begin{table}[htdp]
\caption{Allowed range of $c$ parameters in the SM case. These parameters satisfy  $0<\chi^2<10$ for the SM case. 
The corresponding figure is \ref{quarksm} in Appendix A.}
\begin{center} 
\begin{tabular}{|c|c|c|c|c|c|}
\hline
parameter & range  &parameter & range&parameter & range \\
\hline
\hline
$c_{Q_1}$ & [0,3.0] &$c_{D_1}$ &[0.78,4]&$c_{U_1}$&[-0.97,3.98] \\
$c_{Q_2}$ & [-1.95,2.36] &$c_{D_2} $&[0.39,3.02]&$c_{U_2} $&[-1.99,2.43]  \\
$c_{Q_3}$ & [-3,1] &$c_{D_3} $&[0.39,2.21]&$c_{U_3} $&[-4,1.0]  \\
\hline 
\end{tabular}
\end{center}
\label{hadronrangesm}
\end{table}

\subsubsection{SM Leptonic Mass fits}
 Unlike the quark case, the  fits in the leptonic sector are far more difficult and more constraining
 due to the small mass differences and the large mixing in the neutrino sector.  As mentioned,  we will consider two different cases of 
 neutrino masses while fitting the leptonic data. 
 
 \noindent 
(a) LLHH higher dimensional operator \newline
 Planck scale lepton number violation is an interesting idea which manifests itself with higher dimensional operator suppressed by the Planck scale. 
 In four dimensions such an operator generates too small neutrino mases. It is typically used as a perturbation over an existing neutrino mass model \cite{umashankar}. 
 If not,  it needs an enhancement of $\mathcal{O}(10^3-10^4)$ to be consistent with the data. In the standard RS
  framework close to the weak scale with bulk fermions, this higher dimensional operator is still constrained however for different reasons. 
  While the neutrino masses can be fit by placing the doublet fields $L$ close to the UV brane, the charged lepton masses become very tiny unless
  the singlet fields (E) are placed deep in the IR\cite{iyer}. This leads to inconsistencies in the theory with large non-perturbative Yukawa
  couplings. The question arises whether the situation repeats itself when we consider the modified RS setup.
This can be checked as follows. 
 The neutrino masses are generated by the higher dimensional operator as given  in Eq.(\ref{smaction}). 
 The corresponding neutrino mass matrix is given by Eq.(\ref{smneutrino1})
 while the mass matrix for the charged leptons is given by Eq.(\ref{SMmasses}). 

For simplicity assume $c_{L_i} =c_L \forall$ i. For $c_L<0.5$ the mass matrix in Eq.(\ref{smneutrino1}) becomes
\begin{equation}
  m_\nu = \kappa' \frac{ v^2sin^2(\beta)}{2\epsilon\Lambda}(1-2c_L)
\label{largeneutrino}
\end{equation}
It is clear that $c_L\sim -4$ is required to get neutrino masses $\mathcal{O}(0.04)$ eV for a warp factor for $\epsilon \sim 10^{-2}$.
As $c_L$ increases, beyond 0.5, this formula is no longer valid,  the neutrino masses become smaller and hence do not fit 
the neutrino mass data with $\mathcal{O}(1)$ Yukawa couplings. Thus a mildly negative $c_L$ should be able to fit the data without large inconsistencies.

A second enhancement can also come from the $\kappa'$, which is the corresponding $\mathcal{O}(1)$ Yukawa. With this in mind, we enhance the range 
of the scanning of the Yukawa couplings from $0.08$ to $4$ to $0.08$ to $10$. This would help us to accommodate $c_{L}$ values close to $\sim -1$. 
The final scanning ranges we have chosen are:
the doublets ($c_{L_i}$) are varied  between -1.5 and 0.5,  while the charged singlets were scanned between 0 and 4. The region of $c$ values  which give
a good fit to leptonic masses, \textit{i.e,}  satisfying the constraint $0<\chi^2<10$, is presented in Table[\ref{leptonrangeLLHHsm}]. The plots for these ranges
of  $c$ values are presented in Figs.(\ref{leptonllhhsm}) in Appendix A.

\begin{table}[htdp]
\caption{Ranges for scanned regions of the  bulk leptonic parameters for the LLHH in the SM case which satisfy $0<\chi^2<10$.}
\begin{center} 
\begin{tabular}{|c|c|c|c|c|c|}
\hline
parameter & range  &parameter & range \\
\hline
\hline
$c_{L_1}$ &[ -1.5,-1.15] &$c_{E_1}$ &[2.8,4.0] \\
$c_{L_2}$ & [-1.5,-0.97] &$c_{E_2} $&[1.8,2.4] \\
$c_{L_3}$ & [-1.5,-1.22] &$c_{E_3} $&[1.2,1.69]  \\
\hline 
\end{tabular}
\end{center}
\label{leptonrangeLLHHsm}
\end{table}

\noindent
(b)Dirac type Neutrinos\newline
The case of Dirac neutrinos  is interesting possibility though it requires imposition of a global lepton number conservation\footnote{In fact, it is possible to
hide lepton number violation in this case through a careful location of the right handed fermion fields \cite{planckgher}. We will not consider this case here.}. 
The running of the neutrino masses from the weak scale to high scale is different in this case. However with a normal hierarchy of neutrinos and low tan$\beta$ the differences
are insignificant\cite{xing2,xing3}. 

Assuming that there is not
much of a difference for normal hierarchy, we choose the following scanning range for the $c$ parameters. 
The  doublets ($c_{L_i}$)  and charged lepton singlets ($c_{E_i}$) are scanned within  the range -1 to 4.5, while 
the neutrino singlets were scanned in the range 3.5 to 9. Such a large value of the bulk mass parameters for 
the singlets is needed to suppress the corresponding neutrino 
masses sufficiently. The  $\mathcal{O}$(1) Yukawa parameters were varied between 0.08 and 4.
 Comparing the results of Dirac neutrino mass fits with that  of the weak scale RS models,\cite{iyer}, we find that the $c_N$ are roughly
 a factor $7-8$ larger  compared to the $c_{N}$ at the weak scale. This is purely because  of the weaker warp factor we are considering
 in the present case. Increasing the range of the $O(1)$ Yukawa parameters would only make things worse.  
 The ranges for the $c$ values corresponding to SM fits with Dirac neutrinos case are presented Table[\ref{leptonrangediracsm}]. 
The plots for the $c$ parameters are presented  case in Fig[\ref{leptondiracsm}] in Appendix A. 

\begin{table}[htbp]
\caption{Ranges for the scanned regions of the  bulk leptonic parameters for the Dirac case which satisfy  $0<\chi^2<10$ for the SM case.}
\begin{center} 
\begin{tabular}{|c|c|c|c|c|c|}
\hline
parameter & range  &parameter & range&parameter & range \\
\hline
\hline
$c_{L_1}$ &[ -1,2.9] &$c_{E_1}$ &[0.39,3.62]&$c_{N_1}$&[5.29,8.97] \\
$c_{L_2}$ & [-0.99,2.7] &$c_{E_2} $&[-1.0,2.63]&$c_{N_2} $&[5.31,8.99] \\
$c_{L_3}$ & [-0.99,1.98] &$c_{E_3} $&[-0.99,1.93]&$c_{N_3} $&[5.12,8.97]  \\
\hline 
\end{tabular}
\end{center}
\label{leptonrangediracsm}
\end{table}

\subsection{Supersymmetric Case}

The analysis for the case with bulk supersymmetry is similar to the SM case.  The GUT scale values are derived using the
supersymmetric RGE at the two loop instead of the SM ones. For the neutrinos however, one loop RGE were used  with  experimental inputs at the weak scale. The running of the masses
are not dependent on the mixing angles for a low tan$\beta$. Supersymmetry threshold corrections can play an important role while
deriving the running masses.  Running masses in the supersymmetric framework were obtained  using the relevant 
matching conditions. As is well known, these effects are significant at large tan$\beta$ and the corrections to the neutrino
running  through $Y_D$ and $Y_E$ were considered \cite{Antusch}.

The GUT scale masses and mixings chosen for the scan corresponded to tan$\beta=10$
and are given in Table[\ref{inputtable3}] and [\ref{inputtable4}]. The results of the the scan \textit{i.e,} the ranges 
for the $c$ parameters are weakly dependent on tan$\beta$ 
and can be applied for studying phenomenology for up to tan$\beta\sim 25$.
\begin{table}[htbp]
\caption{GUT scale Masses with supersymmetry for tan$\beta=10$}
\begin{center}
\begin{tabular}{|c|c|c|c|}
\hline 
Mass &Mass&Mass&Mass squared Differences\\
(MeV )& (GeV)& MeV&$eV^2$ \\
\hline
\hline
$m_u = 0.49^{+0.20}_{-0.17}$&$m_c=0.236^{+0.037}_{-0.036}$&$m_e = 0.28^{+0.0000007}_{-0.0000007}$&$\Delta m^2_{12} = 1.6  ^{+0.20}_{-0.21} \times 10^{-4} $\\
$m_{d} = 0.70^{+0.31}_{-0.31}$&$m_b=0.79^{+0.04}_{-0.04}$&$m_{\mu} = 59.9^{+0.000005}_{-0.000005}$&$\Delta m_{23}^2 = 3.2^{+0.13}_{-0.13}  \times 10^{-3}    $\\
$m_{s} = 13^{+4}_{-0.4}$&$m_t=92.2^{+9.6}_{-7.8}$&$m_{\tau} = 1021^{+0.1}_{-0.1}$&-\\
\hline
\end{tabular}
\end{center}
\label{inputtable3}
\end{table}

\begin{table}[htbp]
\caption{Mixing angles for the quarks and leptons at GUT scale with supersymmetry for tan$\beta=10$ }
\begin{center}
\begin{tabular}{|c|c|c|}
\hline 
mixing angles(CKM)&Mixing angles (PMNS)\\
\hline
\hline
$ \theta_{12} =0.226^{+0.00087}_{-0.00087}$ &$ \theta_{12} =0.59^{+0.02}_{-0.015}$ \\ 
$\theta_{23} = 0.0415^{+0.00019}_{-0.00019}$&$\theta_{23} = 0.79^{+0.12}_{-0.12}$\\
$\theta_{13}=0.0035^{+0.001}_{-0.001}$&$\theta_{13}=0.154^{+0.016}_{-0.016}$\\
\hline
\end{tabular}
\end{center}
\label{inputtable4}
\end{table}

\subsubsection{Quark Case}
The range chosen for the scan are the same as that for the SM case i.e.  $-2<c_{Q_1,Q_2}<4$, $-3<c_{Q_3}<1$ for 
the doublets. $-2<c_{d_1,d_2,d_3}<3.5$, for the down type singlets and $-2<c_{u_1,u_2}<4$, $-4<c_{u_3}<1$ for the up type singlets. The regions of $c$ parameter space which satisfy the 
constraint of  $0<\chi^2<10$ for the chosen scanning range are shown in Fig.(\ref{quark}) in Appendix B and the ranges are outlined in Table[\ref{hadronrange}]. 

\begin{table}[htbp]
\caption{Ranges for the scanned regions of bulk hadronic parameters which satisfy $0<\chi^2<10$ for the supersymmetric case. }
\begin{center} 
\begin{tabular}{|c|c|c|c|c|c|}
\hline
parameter & range  &parameter & range&parameter & range \\
\hline
\hline
$c_{Q_1}$ & [-0.16,3.12] &$c_{D_1}$ &[-0.5,4]&$c_{U_1}$&[-1.6,4.0] \\
$c_{Q_2}$ & [-1.32,2.34] &$c_{D_2} $&[-1.9,2.5]&$c_{U_2} $&[-2,2.4]  \\
$c_{Q_3}$ & [-3,1] &$c_{D_3} $&[-2,1.7]&$c_{U_3} $&[-4,1.0]  \\
\hline 
\end{tabular}
\end{center}
\label{hadronrange}
\end{table}

\subsubsection{Leptonic case}
Similar to the SM scenario two cases of neutrino mass generation are considered. The GUT scale input values for the $\chi^2$ is given in Table[\ref{inputtable3}] and [\ref{inputtable4}]. 
\newline

\noindent
(a)LLHH case\newline
The results of the scan of the LLHH case is very similar for both the SM case and the supersymmetric case. The expression for the neutrino mass matrix 
is given in Eq.(\ref{neutrinomassmatrix1}). For the neutrino sector we allow the $\mathcal{O}$(1) Yukawa coupling to vary between -10 and 10 with a minimum of 0.08 while that for
the charged leptons are varied between -4 and 4 with a minimum of 0.08. The doublets were scanned between -1.5 and 0.5 while the charged singlets were scanned between 0 and 4. 
The ranges for the $c$ parameters for the LLHH case for the chosen scanning range satisfying the constraint $0<\chi^2<10$, is presented in 
Table[\ref{leptonrangeLLHH}] and the plots for the $c$ values are presented in Figs.(\ref{leptonllhh}) in Appendix B.

\begin{table}[htdp]
\caption{Ranges for scanned regions of the  bulk leptonic parameters for the LLHH scenario in the supersymmetric case which satisfy $0<\chi^2<10$.}
\begin{center} 
\begin{tabular}{|c|c|c|c|c|c|}
\hline
parameter & range  &parameter & range \\
\hline
\hline
$c_{L_1}$ &[ -1.5,-0.22] &$c_{E_1}$ &[2.6,3.7] \\
$c_{L_2}$ & [-1.5,0.08] &$c_{E_2} $&[2.0,2.57] \\
$c_{L_3}$ & [-1.5,0.04] &$c_{E_3} $&[1.1,1.8]  \\
\hline 
\end{tabular}
\end{center}
\label{leptonrangeLLHH}
\end{table}

\noindent
(b)Dirac Neutrinos \newline
The expression for the mass matrix for the all the leptons is given by Eq.(\ref{chargedfermionIR})
The scanning range for the $c$ values of all the doublets and charged lepton singlets was in the range -1 to 4.5, while 
the neutrino singlets were scanned in the range 3.5 to 9. The magnitude of
$\mathcal{O}$(1) Yukawa parameters were varied between 0.08 and 4. The regions of the $c$ parameters satisfying the constraint $0<\chi^2<10$ for the scanned ranges are presented in Table[\ref{leptonrangedirac}].  
The ranges are presented in Fig.(\ref{leptondirac}) in Appendix B.

\begin{table}[htdp]
\caption{Ranges for the scanned regions of the  bulk leptonic parameters for the Dirac case with supersymmetry which satisfy  $0<\chi^2<10$ for the supersymmetric case.}
\begin{center} 
\begin{tabular}{|c|c|c|c|c|c|}
\hline
parameter & range  &parameter & range&parameter & range \\
\hline
\hline
$c_{L_1}$ &[ -1,2.6] &$c_{E_1}$ &[-0.86,3.46]&$c_{N_1}$&[5.68,8.9] \\
$c_{L_2}$ & [-0.99,2.21] &$c_{E_2} $&[-1,2.24]&$c_{N_2} $&[5.67,8.99] \\
$c_{L_3}$ & [-1,1.54] &$c_{E_3} $&[-1,1.49]&$c_{N_3} $&[5.64,8.99]  \\
\hline 
\end{tabular}
\end{center}
\label{leptonrangedirac}
\end{table}

To summarize, on comparing the SM and the SUSY fits, we find that within a given generation, the fields have a tendency to be localized slightly towards the IR for the SUSY case
than for the SM case. This effect is more pronounced in the down sector and increases with tan$\beta$. A comparison between the fits for the SM case and the SUSY case for 
$tan\beta=10$ and $50$ are presented in Figs.[\ref{comparative}] in Appendix C.
The underlying features of the fit in which the first two generations including the neutrinos
are elementary from the ADS/CFT point of view while the third generation fermions $(t_L,t_R)$ having a tendency to be partially composite or composite, is maintained for both the SM and
the SUSY case.
   From the choice of the $c$ parameters, we find that the LLHH case admits a better fit to the neutrino data than the Dirac case. This is contrary
   to the observations made in \cite{iyer} in normal RS where the $c$ parameters for all the leptons were close to unity. It
   thus offered a more viable alternative than the LLHH case. In order to compensate for the weak warp factor in the Dirac case the right handed neutrinos had bulk masses $c_N\sim 7$.
   This weak warping however, works in favour of the LLHH case where for $c<0.5$ the effective 4D suppression scale is of the 
   $\mathcal{O}$($M_{GUT}$) resulting in fits with $c$ parameters closer to unity.

\section{SUSY Spectrum and Flavour Phenomenology}

There are several ways to break supersymmetry within this RS set up at the GUT scale (see for example discussion in \cite{choi1,choi2, Dudas,Brummer,nomura}.
In the present work, we will consider only one particular set up which manifestly demonstrates the flavour structure  of the fermions within the soft terms. 
More detailed analysis of supersymmetric spectrum will be addressed in \cite{iyer2}. We will assume in the following that supersymmetric breaking happens
on the IR brane, or the GUT brane. Unlike the work of \cite{Hallnomura} and \cite{Brummer} we will not arrange the SM fields in any particular GUT representation.
As has been discussed in these works, a GUT structure can be arranged with possible solutions for proton decay and doublet triplet splitting.
Instead we parameterize SUSY breaking in terms of a single four dimensional  spurion chiral superfield, $X = \theta^2 F$, which is localized on 
the GUT brane. However, for soft masses generated at the Planck brane as in \cite{Dudas}, one may then impose the GS anomaly cancellation conditions on bulk masses to ensure unification of couplings
at the Planck scale. We do not impose any such conditions as the soft masses are generated at the GUT scale.

In the limit, the higher KK modes are decoupled from the GUT scale physics \cite{Marti},  the K{\"a}hler potential relevant for the scalar mass terms is given by 

\begin{multline}
\mathcal{K}^{(4)}= \int dy \delta(y-\pi R )e^{-2 k \pi R}k^{-2}X^\dagger X\left(
    \beta_{q,ij}Q^\dagger_iQ_j
 +  \beta_{u,ij}U^\dagger_iU_j
+  \beta_{d,ij}D^\dagger_iD_j+ \gamma_{u,d}H_{u,d}^\dagger H_{u,d} +  \ldots \right) 
\label{kahler}
\end{multline}
where  $\beta$ have dimensional carrying negative mass dimensions of -1 ( as the matter fields are five dimensional).  $\gamma_{u,d}$ are $\mathcal{O}(1)$
parameters. 

The sfermion mass matrix  is generated when the $X$ fields get a vacuum expectation value  $ m_{\tilde{f}}^2 \sim k^{-2} <X>^\dagger< X> Q^\dagger Q$. 
The mass matrix will however not be diagonal in flavour space.  In the canonical basis,  (\ref{kahler}), the mass matrices take the form 
\begin{equation}
 (m_{\tilde{f}}^2)_{ij} = m_{3/2}^2~\hat \beta_{ij}~e^{(1-c_i-c_j)k R \pi}\xi(c_i)\xi(c_j)
\label{softmassmatrix2}
\end{equation} 
where $\hat \beta_{ij}=2k \beta_{ij}$ are dimensionless $\mathcal{O}(1)$ parameters. $\xi(c_i)$ are defined in Eq.(\ref{SMmasses}).  And the gravitino
mass is defined as 
\begin{equation}
m_{3/2}^2 = { <F>^2 \over k^2 } = { <F>^2 \over M_{Pl}^2 }
\end{equation} 
The Higgs fields are localised on the GUT brane, their masses are given by   $m^2_{H_u,H_d} =  \gamma_{u,d} ~m_{3/2}^2$.

The A-terms are generated from the higher dimensional operators in the super potential of the type :
\begin{equation}
 W^{(4)} = \int dy \delta(y-\pi R) e^{-3ky}k^{-1}X\left( \tilde{A}^u_{ij} H_uQ_iu_j+\tilde{A}^d_{ij} H_dQ_id_j+\tilde{A}^e_{ij} H_dL_iE_j + \ldots \right)
\label{Aterms}
 \end{equation}
where the $\tilde{A}$ are dimensionful parameters having mass dimension -1.  Substituting for the \textit{vev} of the $X$, we have for the
four dimensional trilinear couplings  at the GUT scale: 
\begin{equation}
 A^{u,d}_{ij}=m_{3/2} A'_{ij}e^{(1-c_{i}-c'_{j})kR\pi}\xi(c_i)\xi(c'_j)
\label{Aterms1}
\end{equation}
where we defined the dimensionless $\mathcal{O}(1)$ parameters as $ A'=2k \tilde{A}$. The structure of the A terms and the corresponding 
fermion  mass matrix are similar and they differ only
by the choice of the $\mathcal{O}$(1) parameters. Choosing $A'=2kY'$, makes the down sector A terms diagonal in the mass basis of the fermions at 
the GUT scale. Henceforth, we shall work in this
basis, with the $\mathcal{O}$(1) parameters of the A terms proportional to the $\mathcal{O}$(1) Yukawa parameters. 

The masses for the gauginos are obtained from the following operator in the lagrangian
\begin{equation}
\mathcal{L}=\int d^2\theta k^{-1}X\mathcal{W}_{A\alpha}\mathcal{W}^{\alpha}_A
\end{equation}
At $M_{GUT}$ their masses will be be $m_{1/2}=f m_{3/2}$ where
 $f$ is a $\mathcal{O}(1)$ parameter. $m_{1/2}$ will be treated as an independent parameter. They
are independent of the position of localization of X. as the profile for the gauginos is flat corresponding
to  a bulk mass parameter of 0.5 \cite{Marti,Gherghetta1}.

While the above equations set the boundary conditions at the high scale, the weak scale spectrum is determined by the
RGE evolution. In the present case, the spectrum at the high scale is completely non-universal as determined by the 
profiles of the zero modes of the matter chiral superfields. The structure of soft terms discussed here is similar to the ideas of
flavourful supersymmetry discussed by \cite{Nomura1,Nomura2} and more recently by \cite{Ramond}. In the following we will present two example points one for the
LHLH higher dimensional operator case and another for the Dirac case.

 To begin with, in both the examples, 
we consider that all the $\mathcal{O} (1)$ parameters appearing in the definitions of the soft parameters are proportional
to the unit matrix. We will explicitly mention any deviations as required by the phenomenology when presenting numerical examples.
This would mean that the matrices, $A', \hat{\beta} $ in Eqs. (\ref{softmassmatrix2}, \ref{Aterms1}) are
proportional to unit matrix and the parameters $\gamma_u, \gamma_d$ in Eq.(\ref{kahler}) are equal to one.  However, as we
will see below, they play an important role in low energy phenomenology and one might frequently require to vary them within
the $\mathcal{O}(1)$ range, to satisfy phenomenological constraints.  While  studying the flavour phenomenology, we make 
sure that the soft terms are present in the super-CKM basis. The low-energy spectrum has been computed numerically 
using the spectrum generator SUSEFLAV \cite{suseflav}. 

\subsubsection{LHLH operator case} 

In this case we consider the  following point, (\ref{samplepointllhh}), in the $c$ parameter space. It has a $\chi^2$ of  $5.5$ for the
hadronic sector and $0.7341$ for the leptonic sector.  
As expected it has a mostly composite right handed top quark. In addition, the leptonic doublets are also 
significantly composite in this case. 

\begin{eqnarray}
 c_{Q_1}=2.740\hspace{.5cm}c_{D_1}=0.722\hspace{.5cm}c_{U_1}=0.4024\hspace{0.5cm}c_{L_1}=-1.497\hspace{0.5cm}c_{E_1}=3.634\nonumber\\
c_{Q_2}=1.920\hspace{.5cm}c_{D_2}=0.729\hspace{.5cm}c_{U_2}=0.0652\hspace{0.5cm}c_{L_2}=-0.224\hspace{0.5cm}c_{E_2}=2.290\nonumber\\
 c_{Q_3}=0.960\hspace{.5cm}c_{D_3}=0.801\hspace{.5cm}c_{U_3}=-3.5615\hspace{0.5cm}c_{L_3}=-1.0738\hspace{0.5cm}c_{E_3}=1.769
 \label{samplepointllhh}
 \end{eqnarray}

The choice of $\mathcal{O}$(1) parameters
in the soft sector plays a role in determining the nature of the low energy spectrum.  
For a given set of $c$ parameters, a naive choice
of one for all the $\mathcal{O}$(1) parameters in the soft sector may or may not lead to an acceptable spectrum
 at $M_{susy}$. 
For the LLHH case corresponding to the choice in Eq.(\ref{samplepointllhh}) the $\mathcal{O}$(1) parameters for all 
the soft masses are taken to be 1. The $\mathcal{O}$(1) parameters for the A terms are chosen to be
$\hat A^{u}=1.02Y^{'u}$ while  
$\hat A^{u}=Y^{'d}$ and $\hat A^{e}=0.6Y^{'e}$. Corresponding to these choices of the $\mathcal{O}$(1) parameters and the $c$ values in Eq.(\ref{samplepointllhh}), the soft breaking terms
at the GUT scale in $GeV$ are given as:
{\small{
\begin{equation*}
m_Q  =\begin{bmatrix}
       0.001&  -0.03&  -0.27\\
  -0.03  & 0.85  & 7.6\\
  -0.27  & 7.6 &  68.7
     \end{bmatrix}; m_U  = \begin{bmatrix}
        11.5 & -63.5  & 156.1\\
  -63.5  & 349.6 & -859.1\\
   156.1  &-859.1 &  2110.8
     \end{bmatrix};
          m_D  = \begin{bmatrix}
              105.03&  -90.4&   155.2\\
  -90.4 &  77.8 & -133.6\\
   155.2  &-133.6  & 229.5
            \end{bmatrix}
            \end{equation*}
            \begin{equation*}
              A_U=\begin{bmatrix}
                                               -0.002	&-0.09&	-0.84\\	
                                               -0.0002	&1.09&	-6.3\\	
                                               -10^{-6}&	0.01&	439.4
                                              \end{bmatrix}; 
      A_D=\begin{bmatrix}
           0.03&0&0\\
           0&-0.40&0\\
           0&0&	-40.6
          \end{bmatrix};            m_L=\begin{bmatrix}
                                       7.7 &   8.8&  -147.5\\
                                      8.8  & 9.9 & -167.0\\
                                      -147.5 & -167.0 &  2798.4
                                                             \end{bmatrix}
            \end{equation*}

           \begin{equation}
             m_E  =\begin{bmatrix}
               0.001&  -0.017&   0.05\\
                 -0.017&   0.26&  -0.89\\
                 0.05&  -0.89&   3.04
                    \end{bmatrix};         A_E=   \begin{bmatrix}   
                                                             0.008&0&0\\
                                                             0&1.81&0\\
                                                             0&0&-31.3
                                                            \end{bmatrix}
                                                                         \end{equation}}}

A couple of interesting features of the above spectrum are (i) at least one of the soft masses  is tachyonic (ii) significant amount
of flavour violation present at the high scale. However at the weak scale, things are significantly different. 
This is because the RG running is quite different for the diagonal terms compared to the off-diagonal ones. In fact, the 
off-diagonal entries barely run, where as the corrections to the diagonal ones are quite significant.  
 As an illustration, consider the slepton mass matrix at the weak scale. The analytic output at the weak scale for the diagonal terms can be approximated as
 \begin{eqnarray}
   \tilde M^2_{L_{1,2}} \simeq m_{L_{1,2}}^2 + 0.5M_{1/2}^2 \\ \nonumber
     \tilde M^2_{L_3} \simeq m_{L_3}^2 + 0.5M_{1/2}^2\\ \nonumber
	\tilde M^2_{E_{1,2}} \simeq m_{E_{1,2}}^2 + 0.15M_{1/2}^2 \\ \nonumber
	\tilde M^2_{E_3} \simeq m_{E_3}^2 + 0.15M_{1/2}^2\\ \nonumber
 \end{eqnarray}
which receive gauge contributions while the off diagonal elements do not. The A terms are not large enough to make the off diagonal elements of the soft mass matrices comparable
with the diagonal terms. An example for the sleptons for the case under consideration is given as  
\begin{eqnarray}
 m_L^2( m_{susy})&=&\begin{bmatrix}
  2.2\times10^5   & 77.0 &  -2.1\times10^4\\
     77.0  &  2.2\times10^5 &  -2.7\times10^4\\
     -2.1\times10^4  & -2.7\times10^4  &  7.7\times10^6
 \end{bmatrix} \text{GeV}^2 \nonumber \\
 m_E^2(m_{susy})&=&\begin{bmatrix}
                          1.1\times10^{6}  &  3.4\times10^{-4} &   2.2\times10^{-1}\\
                              3.4\times10^{-4}  &  1.1\times10^{6} &   5.9\times10^1\\
                                2.2\times10^{-1}  &  5.9\times10^{1}  &  5.4\times10^5
                                                    \end{bmatrix} \text{GeV}^2
\end{eqnarray}

We see that the off-diagonal entry has barely enhanced where as the diagonal entries have been significantly modified.  
Constraints from flavour violation would restrict the mass scales of $m_{3/2}$ and $M_{1/2}$. The most 
stringent constraints are from the transitions between the first two generations \textit{ie}  from $K^0 \to \bar{K}^0$ and
$\mu \to e + \gamma$. The expressions for the $K_L-K_S$ mass difference and the branching fractions for $\mu \to e + \gamma$  can be found in \cite{Gabbiani,Gabbiani1}.
We impose the flavour constraints from all the existing data on the $\delta$ parameters.

Bounds from the flavour violating processes are obtained  using the mass insertion approximation \cite{Gabbiani} and the results of \cite{vempati}
defined in the Super-CKM basis. The flavour violating indices are defined as
$\delta_{ij} (i \neq j) = { (U^\dagger M^2_{soft} U )_{ij}  \over  m_{susy}} (i \neq j)$ are evaluated in the basis in which the down sector is diagonal, with U being the rotation matrix which rotates the corresponding 
fermion mass matrix. The $\delta$ are evaluated at the weak scale  and we scale the bounds of \cite{vempati} to the present mass scales.
In Table[\ref{softspectrumllhh}] we present the low energy spectrum corresponding to the sample point in Eq.(\ref{samplepointllhh}). The low energy $\delta's$ are 
  presented in Table \ref{deltallhh}.

\begin{table}
\caption{Experimental upper bounds on the $\delta^{down}$ obtained for $\tilde m_q=2.1$ TeV and $\tilde m_l=0.7$ TeV}
 \begin{center}
\begin{tabular}{ |c|p{1.5cm}p{1.0cm}|p{1.5cm}p{1.5cm}|p{1.5cm}p{1.5cm}|p{1.5cm}p{1.5cm}| }
  \cline{1-9}
  \cline{1-9}
 (i,j)&$|\delta^Q_{LL}|$ & $|\delta^L_{LL}|$&$|\delta^D_{LR}|$ & $|\delta^E_{LR}|$&$|\delta^D_{RL}|$ & $|\delta^E_{RL}|$&$|\delta^D_{RR}|$ & $|\delta^E_{RR}|$ \\
 \cline{1-9}
  12&0.053 &$0.0002$& $0.0003$ &$3.8\times 10^{-6}$ &$0.0003$ &$3.8\times10^{-6}$&0.03&0.03\\
 13&$0.34$  &0.14& $0.06$ &$0.03$&0.06&$0.03$&0.26& -\\
 23&0.61 &$0.16$&  $0.01$&0.04&$0.02$&0.04&0.84&-\\  
   \cline{1-9}
     \end{tabular}
 \end{center}
 \label{exptconst}
\end{table}

\begin{table}[htbp]
\caption{Soft spectrum for LLHH case: $m_{susy}=1.06$ TeV, $m_{\tilde g}=2.64$ TeV, $\mu=3.43$TeV, $tan \beta=25$}
\begin{center}
\begin{tabular}{|c|c|c|c|c|c|c|c|c|c|}
\hline
Parameter&Mass(TeV)&Parameter&Mass(TeV)&Parameter&Mass(TeV)&Parameter&Mass(Tev)&Parameter&Mass(TeV)\\  
\hline 

              $\tilde t_1$ &   0.47&$\tilde b_1$  &   1.01& $\tilde\tau_1$ &   0.726&$\tilde\nu_\tau$  &   2.78&$N_1$&0.465\\
               $\tilde t_2$ &   1.05&$\tilde b_2$  &   2.14&$\tilde\tau_2$ &   2.79&$\tilde\nu_\mu$ &   0.483&$N_2$&0.929\\
             $\tilde c_R$ &   2.24&$\tilde s_R$ &   2.40&$\tilde\mu_R$  &   0.478&$\tilde\nu_e$ &   0.469&$N_3$&3.38\\
             $\tilde c_L$ &   2.48& $\tilde s_L$ &   2.48&$\tilde\mu_L$  &   1.05&-&-&$N_4$&3.39\\
                $\tilde u_R$ &   2.24& $\tilde d_R$  &   2.40&$\tilde e_R$   &   0.476&-&-&$C_1$&0.895\\
                $\tilde u_L$ &   2.48&$\tilde d_L$  &   2.48& $\tilde e_L$   &   1.05&-&-&$C_2$&3.43\\
                $m_{A^0}$&3.23&$m_H^\pm$&3.23&$m_h$&0.12186&$m_H$&3.06&-&-\\
 \hline 
\hline 
\end{tabular}
\end{center}
\label{softspectrumllhh}
\end{table}

 \begin{table}[htbp]
 \caption{Low energy $\delta's$ for quarks and leptons corresponding to the points in Eq.(\ref{samplepointllhh}) for the LLHH case evaluated for $\tilde m_q=2.1$TeV and $\tilde m_l=0.7$ TeV}
 \begin{center}
\begin{tabular}{ |c|p{1.5cm}p{1.0cm}|p{1.5cm}p{1.0cm}|p{1.5cm}p{1.0cm}|p{1.5cm}p{1.5cm}p{1.0cm}| }
  \cline{1-10}
 (i,j)&$|\delta^Q_{LL}|$ & $|\delta^L_{LL}|$&$|\delta^D_{LR}|$ & $|\delta^U_{LR}|$&$|\delta^D_{RL}|$ & $|\delta^U_{RL}|$&$|\delta^D_{RR}|$ & $|\delta^E_{RR}|$&$|\delta^U_{RR}|$ \\
 \cline{1-10}
 12&0.0003 &$0.0001$& $10^{-10}$ &$10^{-8}$&$10^{-8}$&$10^{-5}$& $0.001$&$10^{-10}$&0.001\\
 13&$0.01$  &0.04& $10^{-8}$ &$10^{-8}$&$10^{-6}$&$0.002$&$0.005$&$10^{-7}$& 0.01\\
  23&0.05 &$0.05$&  $10^{-6}$&$10^{-5}$&$10^{-5}$&$0.01$&$0.003$&0.0001&0.07\\  
   \cline{1-10}
     \end{tabular}
 \end{center}
 \label{deltallhh}
\end{table}

\subsubsection{Dirac Case}
 
 For the case where neutrinos are of Dirac type the  $c$ parameters in Eq.(\ref{samplepointdirac}) with  $\chi^2$ of $0.3211$ for the
 hadronic sector and $0.1481$ for the leptonic sector were chosen. The $c$  values for the 
 doublets in this case indicate they are predominantly elementary from the CFT point of view especially for the first two generations. The third generation however
 may be partially composite as in this case. 
 
 \begin{eqnarray}
 c_{Q_1}=1.895\hspace{.5cm}c_{D_1}=1.898\hspace{.5cm}c_{U_1}=1.738\hspace{0.5cm}c_{L_1}=1.293\hspace{0.5cm}c_{E_1}=2.480\hspace{0.5cm}c_{N_1}=6.783\nonumber\\
c_{Q_2}=1.467\hspace{.5cm}c_{D_2}=1.271\hspace{.5cm}c_{U_2}=1.124\hspace{0.5cm}c_{L_2}=1.311\hspace{0.5cm}c_{E_2}=1.406\hspace{0.5cm}c_{N_2}=7.346\nonumber\\
 c_{Q_3}=-0.137\hspace{.5cm}c_{D_3}=1.394\hspace{.5cm}c_{U_3}=-0.356\hspace{0.5cm}c_{L_3}=0.260\hspace{0.5cm}c_{E_3}=0.237\hspace{0.5cm}c_{N_1}=7.332
 \label{samplepointdirac}
 \end{eqnarray}
  The generic feature of soft mass matrices, discussed in the LLHH case, of the  spectrum being tachyonic at the high scale and the diagonal terms evolving more than the off 
  diagonal elements apply to this case as well. Corresponding to the c values in Eq.(\ref{samplepointdirac}), the soft masses at the GUT scale have been evaluated for $m^{GUT}_{3/2}=800$ GeV 
  while choosing $M_{1/2}=1200$ GeV for the three gauginos. The  $\mathcal{O}(1)$ parameters corresponding to $(m_Q)_{33}$ and $(m_U)_{33}$ were chosen to be 4 while for the others they
  are set to be 1. 
$A^{'u}_{ij}=1.15Y^{'u} \forall i,j$, for all the A terms of the up sector while for the down sector and the leptons they were set equal to the 
corresponding $\mathcal{O}$(1) Yukawa couplings. The high scale soft breaking matrices in $GeV$ are given in Eq.(\ref{softgutdirac}). 
  In Table[\ref{softspectrumdirac}] we present the low energy spectrum corresponding to the sample point in Eq.(\ref{samplepointdirac}). The low energy $\delta's$ are 
  presented in Table \ref{deltadirac}.

  \begin{equation*}
 m_Q  =\begin{bmatrix} 
  0.60&   1.2&   30.0\\
   1.2&   8.7&  -19.4\\
   30.0&  -19.4&   2560.6
   \end{bmatrix};m_U =\begin{bmatrix}
      0.7&  -3.8&   17.1\\
  -3.8&   32.1&   70.3\\
   17.1&   70.3&   2971.6
   \end{bmatrix};
   m_D  = \begin{bmatrix}
          0.10&  -0.85&  -1.7\\
  -0.85&   7.1&   14.6\\
  -1.7&   14.6&   29.8  
         \end{bmatrix}
         \end{equation*}
         \begin{equation*}
           A_U=\begin{bmatrix}
         10^{-3}&	-0.42&	-5.8\\	
         10^{-3}&1.18&	0.20\\	
         10^{-5}&-0.005&	488.8
         \end{bmatrix};
         A_D=\begin{bmatrix}
                0.03&0&0\\
                0&0.57&0\\
               0&0&-40.6
              \end{bmatrix};    
 m_L=\begin{bmatrix}
    8.7  & 0.9 &  62.5\\
   0.9  & 0.1  & 7.0\\
   62.5  & 7.0 &  446.3
          
      \end{bmatrix}
      \end{equation*}
      \begin{equation}
        m_E  =\begin{bmatrix}
                        0.2&  -0.4 &  10.5\\
                       -0.4 &  0.7  &-17.9\\
                         1.0 & -17.29&   442.7
                    \end{bmatrix}; A_E=\begin{bmatrix}
                                                        -0.01&0&0\\
                                                       0&-3.01&0\\
                                                       0&0&52.1
             \label{softgutdirac}                                           \end{bmatrix}
      \end{equation}

\begin{table}[htbp]
\caption{Soft spectrum for Dirac case: $m_{susy}=1.05$ TeV, $m_{\tilde g}=2.65$ TeV, $\mu=4.32$TeV, $tan \beta=25$}
\begin{center}
\begin{tabular}{|c|c|c|c|c|c|c|c|c|c|}
\hline
Parameter&Mass(TeV)&Parameter&Mass(TeV)&Parameter&Mass(TeV)&Parameter&Mass(Tev)&Parameter&Mass(TeV)\\  
\hline 

              $\tilde t_1$ &   0.702&$\tilde b_1$  &   2.06& $\tilde\tau_1$ &   0.480&$\tilde\nu_\tau$  &   0.570&$N_1$&0.465\\
               $\tilde t_2$ &   2.31&$\tilde b_2$  &   2.32&$\tilde\tau_2$ &   0.802&$\tilde\nu_\mu$ &   0.624&$N_2$&0.928\\
             $\tilde c_R$ &   2.25&$\tilde s_R$ &   2.36&$\tilde\mu_R$  &   0.608&$\tilde\nu_e$ &   0.625&$N_3$&4.26\\
             $\tilde c_L$ &   2.45& $\tilde s_L$ &   2.45&$\tilde\mu_L$  &   0.902&-&-&$N_4$&4.26\\
                $\tilde u_R$ &   2.25& $\tilde d_R$  &   2.36&$\tilde e_R$   &   0.610&-&-&$C_1$&0.894\\
                $\tilde u_L$ &   2.45&$\tilde d_L$  &   2.45& $\tilde e_L$   &   0.903&-&-&$C_2$&4.32\\
	    $m_{A^0}$&4.18&$m_H^{\pm}$&4.18&$m_h$&0.1235&$m_H$&3.96&-&-\\
 \hline 
\hline 
\end{tabular}
\end{center}
\label{softspectrumdirac}
\end{table}

 \begin{table}[htbp]
 \caption{Low energy $\delta's$ for the Dirac Case corresponding to the point in Eq.(\ref{samplepointdirac}) for Dirac case evaluated for $\tilde m_q=2.1$TeV and $\tilde m_l=0.7$ TeV}
 \begin{center}
\begin{tabular}{ |c|p{1.5cm}p{1.0cm}|p{1.5cm}p{1.0cm}|p{1.5cm}p{1.0cm}|p{1.5cm}p{1.5cm}p{1.5cm}| }
  \cline{1-10}
(ij) &$|\delta^Q_{LL}|$ & $|\delta^L_{LL}|$&$|\delta^D_{LR}|$ & $|\delta^U_{LR}|$&$|\delta^D_{RL}|$ & $|\delta^U_{RL}|$&$|\delta^D_{RR}|$ & $|\delta^E_{RR}|$&$|\delta^U_{RR}|$ \\
 \cline{1-10}
  12&0.0003 &$10^{-6}$& $10^{-10}$ &$10^{-8}$ &$10^{-8}$&$10^{-5}$& $10^{-7}$&$10^{-7}$&0.00005\\
 13&$0.01$  &0.007& $10^{-8}$ &$10^{-8}$&$10^{-5}$&$0.002$&$10^{-6}$&$10^{-4}$&0.06 \\
 23&0.06 &$10^{-4}$&  $10^{-6}$&$10^{-5}$&$10^{-5}$&$0.01$&$10^{-4}$&0.0006&0.001\\  
   \cline{1-10}
     \end{tabular}
 \end{center}
 \label{deltadirac}
\end{table}              

\section{Outlook}
The Randall-Sundrum framework is typically considered  to be the geometric avatar of the Froggatt-Nielsen models. 
In the present work, we have considered a warped extra dimension close to the GUT scale.  We fit the quark  masses
and the CKM  mixing angles and determined the range of the $c$ parameters which give a reasonable $\chi^2$ fit. 
The $\mathcal{O}(1)$ parameters associated with the Yukawa couplings have also been varied accordingly.  
Though the top quark Yukawa is smaller at the high scale compared to the weak scale,   it is still large enough that
one requires a large negative bulk mass parameter for the right handed top quark.  For the leptons, we considered
two particular models for neutrino masses (a) with Planck scale lepton number violating operator and (b) Dirac neutrino
masses.  

The results show that there is a significant difference in the RS models at the weak scale and the RS models at the
GUT scale especially if one focuses on the neutrino sector.  In the weak scale models, the Planck scale lepton
number violating higher dimensional operator was very hard to accommodate with perturbative Yukawa couplings
and thus was highly disfavoured.  The Dirac and the Majorana cases were favoured though they were strongly
constrained by the data from flavour violating rare decays. The situation is sort of reversed in the GUT scale 
RS models, though not exactly. The higher dimensional Planck scale operator fits the data very well, where as
the Dirac case requires larger $c$ values, some times with bulk mass parameters almost an order of magnitude
larger than the cut-off scale.  In the hadronic sector, the situation is not so dramatic. The top quark Yukawa though
reduces at the GUT scale compared to the weak scale, still requires that the right handed top to be located close to
the IR, making it a composite as in the weak scale models. 

The supersymmetric version of the same set up is far more interesting as it can have possible observable signatures
at the weak scale. The main difference in fitting of the fermion masses is due to the presence of the additional parameter 
$\tan\beta$ in the supersymmetric case.  The difference in the $c$ values is not very significant for low values of $\tan\beta$. 
At large $\tan\beta$, for a given generation, the zero modes are localized more towards the IR brane as compared to the SM case.
This effect is more pronounced in the down sector as shown in Figures[\ref{comparative}].
 We parameterise SUSY breaking by a single spurion field  localised on the IR brane. 
The resultant soft masses depend on the profiles of  the zero modes of the chiral superfields and contain
flavour violation.  At the weak scale the constraints from first two generation flavour transitions rule out  light spectrum.
Another significant constraint comes the light Higgs mass at 125 GeV.  The trilinear couplings have the same form as the Yukawa
couplings in this model as long as the $\mathcal{O}$(1) parameters associated with both the parameters are taken to
be proportional to each other. If all the $\mathcal{O}(1)$ parameters at  high scale are take to be exactly unity, the weak
scale values of $A_t$ are  small to generate a  125  GeV  light Higgs, for  stops  of masses $\sim 1.5-2$ 
TeV. However, with a  minor variation of the $\mathcal{O}(1)$ parameters for $A_t$ at the high scale,  125 GeV
Higgs is easily possible.  More variations of supersymmetric breaking and the corresponding spectra will be
discussed later \cite{iyer2}.

 \vskip 1 cm
 \noindent
 \textbf{Acknowledgments}\\  
\newline 
We  thank Emilian Dudas for  discussions and collaboration in the initial stages of this work.  We also thank him 
for a  clarification about a point. AI would like to thank CPhT Ecole Polytechnique for hospitality during his stay 
and Gero Von Gersdorff for  useful discussions.  We also thank  Debtosh Chowdhury for useful inputs.  
We thank S. Uma Sankar for a reference and communications. 
SKV acknowledges support from DST Ramanujam fellowship  SR/S2/RJN-25/2008 of Govt. of India. 

\appendix 

\section{Plots for ranges of c parameters for quarks and leptons for SM fits at the GUT scale }
\label{SM}
\subsection{Range of c parameters for the quarks}
\begin{figure}[H]
\begin{tabular}{cc}
\includegraphics[width=0.5\textwidth,angle=0]{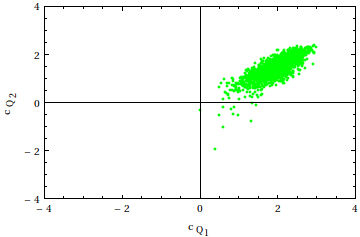} &
\includegraphics[width=0.5\textwidth,angle=0]{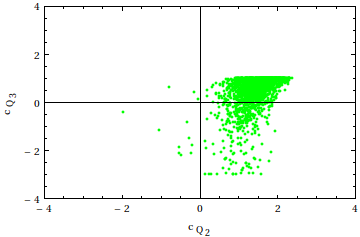} \\
\includegraphics[width=0.5\textwidth,angle=0]{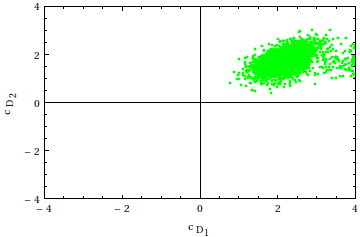} &
\includegraphics[width=0.5\textwidth,angle=0]{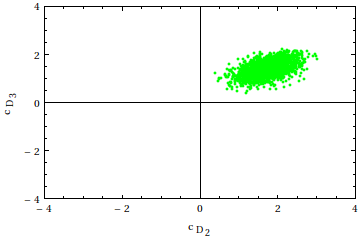} \\
\includegraphics[width=0.5\textwidth,angle=0]{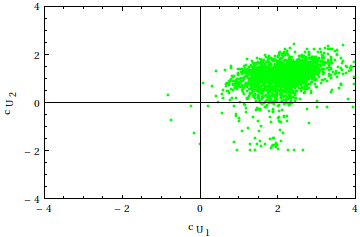} &
 \includegraphics[width=0.5\textwidth,angle=0]{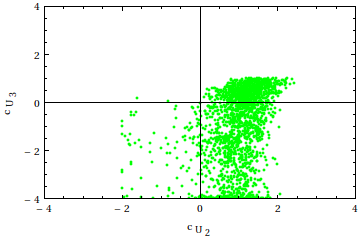} \\
\end{tabular}
 \caption{The points in the above figures correspond to a $\chi^2$ between 1 and 10 for the SM fits at the GUT scale. The plot
represents the parameter space for the bulk masses of the  quarks. 
}
\label{quarksm}
\end{figure}
\newpage
\subsection{Range of c parameters for the for leptons for the LLHH scenario}
\begin{figure}[htbp]
\begin{tabular}{cc}
\includegraphics[width=0.5\textwidth,angle=0]{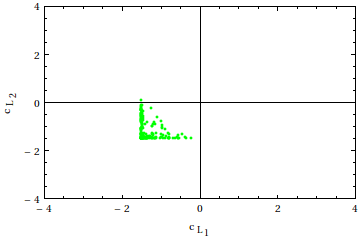} &
\includegraphics[width=0.5\textwidth,angle=0]{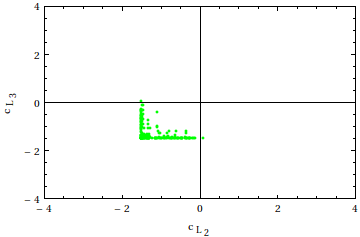} \\
\includegraphics[width=0.5\textwidth,angle=0]{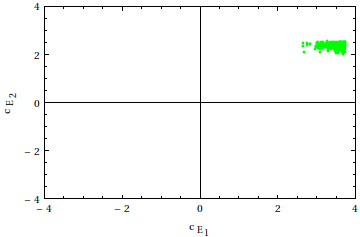} &
\includegraphics[width=0.5\textwidth,angle=0]{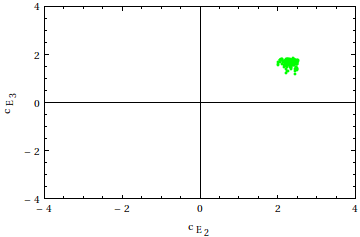} \\
\end{tabular}
 \caption{The points in the above figures correspond to a $\chi^2$ between 1 and 10. The plot
represents the parameter space for the bulk masses of the leptons corresponding to the LLHH case. 
}
\label{leptonllhhsm}
\end{figure}
\newpage
\subsection{Range of c parameters for the leptons.}
\begin{figure}[htbp]
\begin{tabular}{cc}
\includegraphics[width=0.5\textwidth,angle=0]{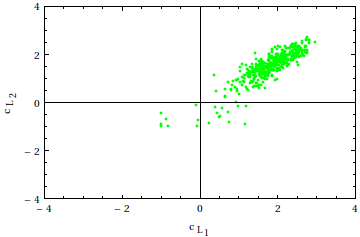} &
\includegraphics[width=0.5\textwidth,angle=0]{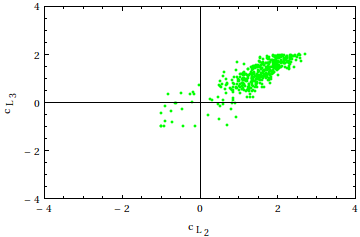} \\
\includegraphics[width=0.5\textwidth,angle=0]{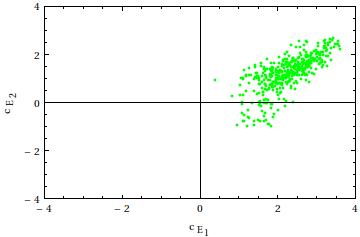} &
\includegraphics[width=0.5\textwidth,angle=0]{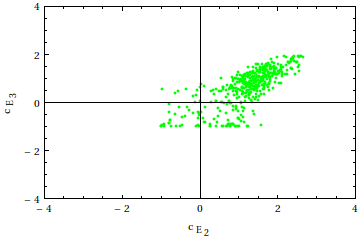} \\
\includegraphics[width=0.5\textwidth,angle=0]{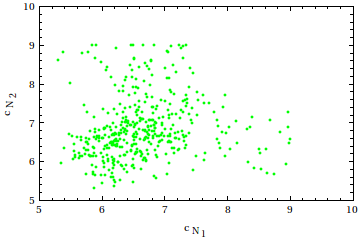} &
\includegraphics[width=0.5\textwidth,angle=0]{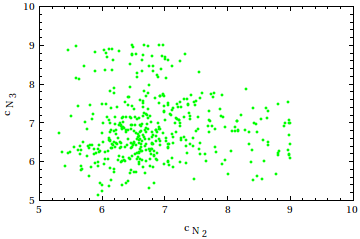} \\
\end{tabular}
 \caption{The points in the above figures correspond to a $\chi^2$ between 1 and 10 for the SM fits at the GUT scale. The plot
represents the parameter space for the bulk masses of the  leptons corresponding to the Dirac type neutrinos. 
}
\label{leptondiracsm}
\end{figure}

\section{Plots for ranges of c parameters for quarks and leptons for the supersymmetric case }
\subsection{Range of c parameters for the quarks}
\begin{figure}[H]
\begin{tabular}{cc}
\includegraphics[width=0.5\textwidth,angle=0]{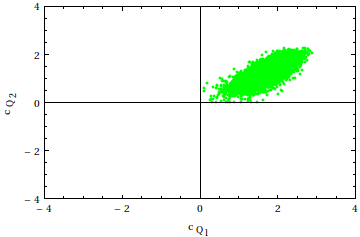} &
\includegraphics[width=0.5\textwidth,angle=0]{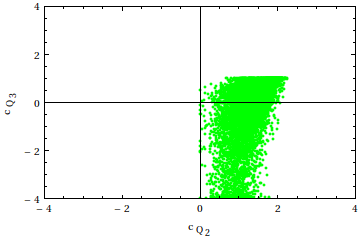} \\
\includegraphics[width=0.5\textwidth,angle=0]{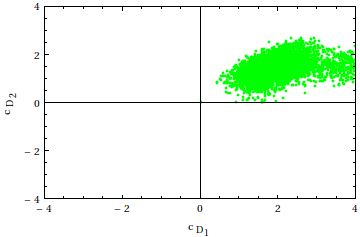} &
\includegraphics[width=0.5\textwidth,angle=0]{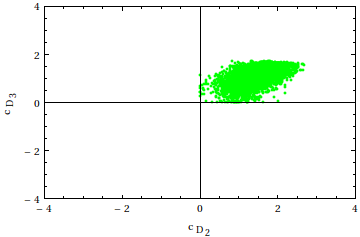} \\
\includegraphics[width=0.5\textwidth,angle=0]{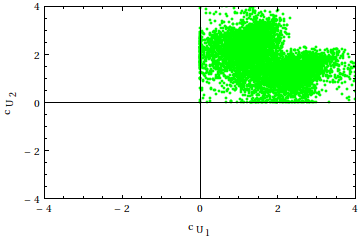} &
 \includegraphics[width=0.5\textwidth,angle=0]{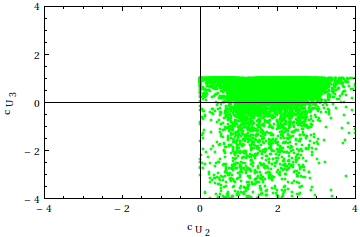} \\
\end{tabular}
 \caption{The points in the above figures correspond to a $\chi^2$ between 1 and 10. The plot
represents the parameter space for the bulk masses of the  quarks. 
}
\label{quark}
\end{figure}
\newpage
\subsection{Range of c parameters for the for leptons for the LLHH scenario}
\begin{figure}[htbp]
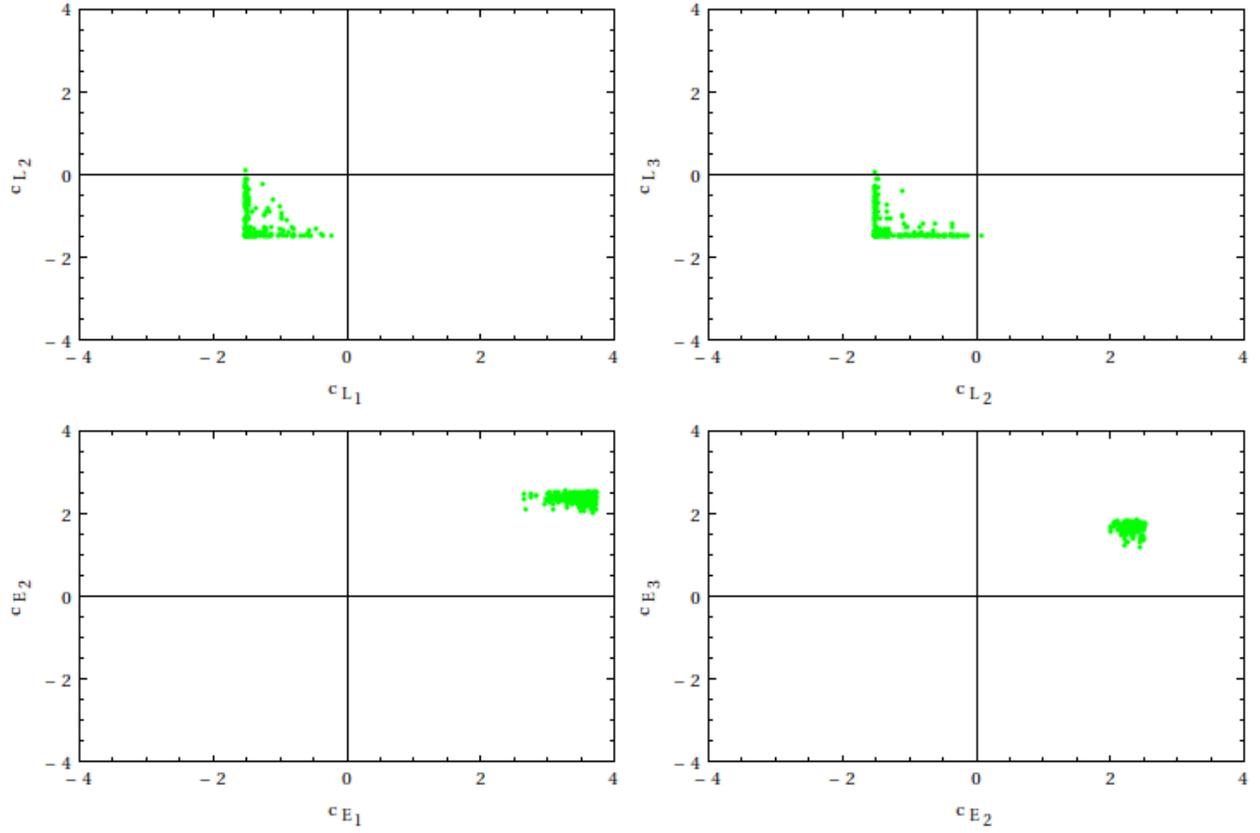

\begin{tabular}{cc}
\includegraphics[width=0.5\textwidth,angle=0]{llhh1.png} &
\includegraphics[width=0.5\textwidth,angle=0]{llhh2.png} \\
\includegraphics[width=0.5\textwidth,angle=0]{llhh3.png} &
\includegraphics[width=0.5\textwidth,angle=0]{llhh4.png} \\
\end{tabular}
 \caption{The points in the above figures correspond to a $\chi^2$ between 1 and 10. The plot
represents the parameter space for the bulk masses of the leptons corresponding to the LLHH case. 
}
\label{leptonllhh}
\end{figure}
\newpage
\subsection{Range of c parameters for the for leptons for the case of Dirac neutrinos.}

\begin{figure}[htbp]
\begin{tabular}{cc}
\includegraphics[width=0.5\textwidth,angle=0]{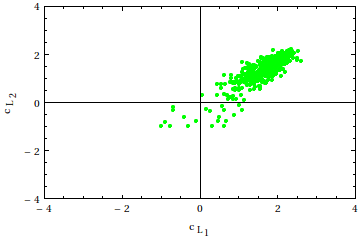} &
\includegraphics[width=0.5\textwidth,angle=0]{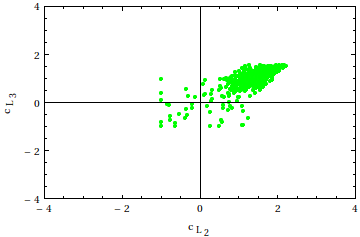} \\
\includegraphics[width=0.5\textwidth,angle=0]{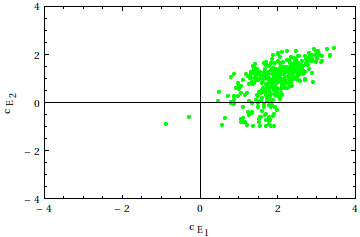} &
\includegraphics[width=0.5\textwidth,angle=0]{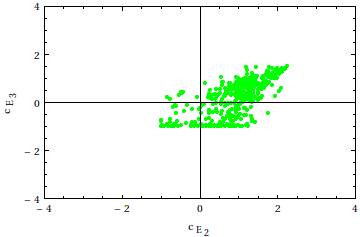} \\
\includegraphics[width=0.5\textwidth,angle=0]{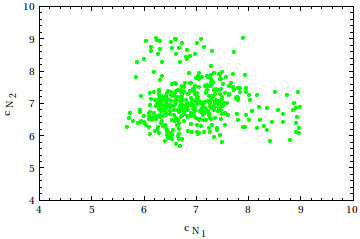} &
\includegraphics[width=0.5\textwidth,angle=0]{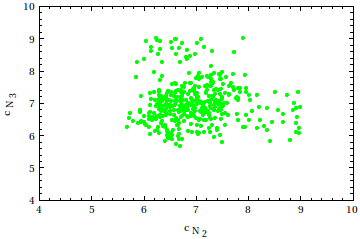} \\
\end{tabular}
 \caption{The points in the above figures correspond to a $\chi^2$ between 1 and 10. The plot
represents the parameter space for the bulk masses of the  leptons corresponding to the Dirac type neutrinos. 
}
\label{leptondirac}
\end{figure}
\newpage
\section{Comparative plot between SM and supersymmetric fits for fermions}
\label{compplota}

\begin{figure}[htbp]
\begin{tabular}{cc}
\includegraphics[width=0.5\textwidth,angle=0]{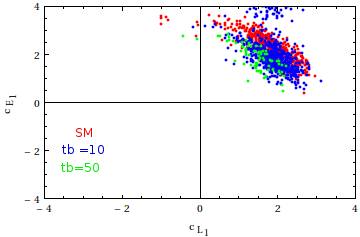} &
\includegraphics[width=0.5\textwidth,angle=0]{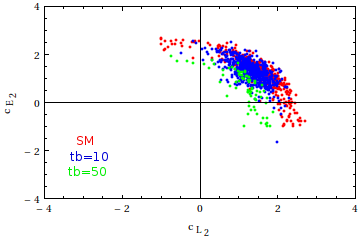} \\
\includegraphics[width=0.5\textwidth,angle=0]{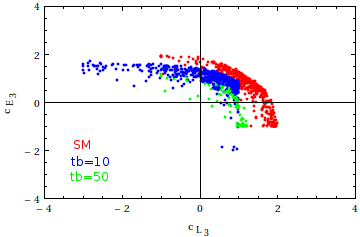} &
\includegraphics[width=0.5\textwidth,angle=0]{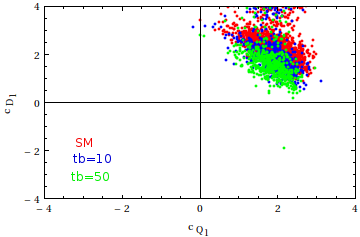} \\
\includegraphics[width=0.5\textwidth,angle=0]{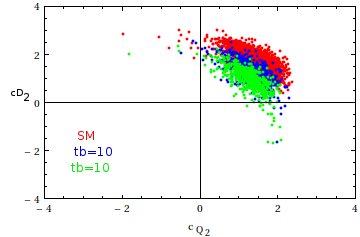} &
\includegraphics[width=0.5\textwidth,angle=0]{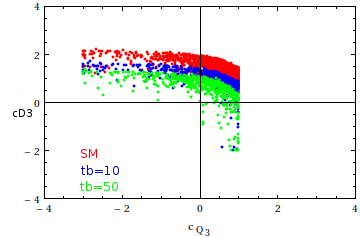} \\
\end{tabular}
 \caption{The plots correspond to comparison between SM and supersymmetric fits for down sector fermions. \textcolor{red}{Red} corresponds to SM, while
 \textcolor{blue}{Blue} and \textcolor{green}{Green} corresponds to supersymmetric case for tan$\beta=10$ and tan$\beta=50$ respectively
}
\label{comparative}
\end{figure}

\bibliographystyle{ieeetr}

       \bibliography{RS.bib}

\end{document}